\documentclass[aps,prb,twocolumn,groupedaddress]{revtex4}

\usepackage{graphicx}

\begin{document}


\title{Origin of positive magnetoresistance in small-amplitude unidirectional lateral superlattices}


\author{Akira Endo}
\email[]{akrendo@issp.u-tokyo.ac.jp}
\author{Yasuhiro Iye}
\affiliation{Institute for Solid State Physics, University of Tokyo, 5-1-5 Kashiwanoha, Kashiwa, Chiba 277-8581, Japan}


\date{\today}

\begin{abstract}
We report quantitative analysis of positive magnetoresistance (PMR) for unidirectional-lateral-superlattice samples with relatively small periods ($a$=92--184 nm) and modulation amplitudes ($V_0$=0.015--0.25 meV). By comparing observed PMR's with ones calculated using experimentally obtained mobilities, quantum mobilities, and $V_0$'s, it is shown that contribution from streaming orbits (SO) accounts for only small fraction of the total PMR\@. For small $V_0$, the limiting magnetic field $B_\mathrm{e}$ of SO can be identified as an inflection point of the magnetoresistance trace. The major part of PMR is ascribed to drift velocity arising from incompleted cyclotron orbits obstructed by scatterings.
\end{abstract}

\pacs{73.23.-b, 73.23.Ad, 75.47.Jn, 73.40.-c}

\maketitle

\section{Introduction\label{Intro}}
Large mean free path ($L$$\gg$1 $\mu$m) of GaAs/AlGaAs-based two-dimensional electron gas (2DEG) and modern nano-fabrication technologies have enabled us to design and fabricate 2DEG samples artificially modulated with length scales much smaller than $L$. The samples have been extensively utilized for experimental investigations of novel physical phenomena that take place in the new artificial environments. \cite{BeenakkerR91} Unidirectional lateral superlattice (ULSL) represents a prototypical and probably the simplest example of such samples; there, a new length scale, the period $a$, and a new energy scale, the amplitude $V_0$, of the periodic potential modulation are introduced to 2DEG\@. These artificial parameters give rise to a number of interesting phenomena through their interplay with parameters inherent in 2DEG, especially when subjected to a perpendicular magnetic field $B$. Magnetotransport reveals intriguing characteristics over the whole span of magnetic field, ranging from low field regime dominated by semiclassical motion of electrons, \cite{Weiss89,Winkler89,Beton90P,Geim92} through quantum Hall regime where several Landau levels are occupied, \cite{Muller95,Tornow96,Milton00,Endo02f,Endo04EP} up to the highest field where only the lowest Landau level is partially occupied \cite{Smet99c,Willett99c,Endo01c}; in the last regime, semiclassical picture is restored with composite Fermions (CFs) taking the place of electrons. Of these magnetotransport features, two observed in low fields, namely, positive magnetoresistance \cite{Beton90P} (PMR) around zero magnetic field and commensurability oscillation \cite{Weiss89,Winkler89} (CO) originating from geometric resonance between the period $a$ and the cyclotron radius $R_\mathrm{c}$=$\hbar k_\mathrm{F}/e|B|$, where $k_\mathrm{F}$=$\sqrt{2 \pi n_e}$ represents the Fermi wave number with $n_e$ the electron density, have the longest history of being studied and are probably the best-known.

The PMR has been ascribed to channeled orbit, or streaming orbit (SO), in which electrons travel along the direction parallel to the modulation ($y$-direction), being confined in a single valley of the periodic potential. \cite{Beton90P} Electrons that happen to have the momentum perpendicular to the modulation ($x$-direction) insufficient to overcome the potential hill constitute SO\@. In a magnetic field $B$, Lorentz force partially cancels the electric force deriving from the confining potential. Therefore the number of SO's decreases with increasing $B$ and finally disappears at the limiting field where Lorentz force balances with the maximum slope of the potential. The \textit{extinction field} $B_\mathrm{e}$ depends on the amplitude and the shape of the potential modulation, and for sinusoidal modulation $V_0 \cos(2\pi x/a)$,
\begin{equation}
B_\mathrm{e}=\frac{2\pi m^* V_0}{ae\hbar k_\mathrm{F}},\label{Be}
\end{equation}
where $m^*$ represents the effective mass of electrons. It follows then that $V_0$ can be deduced from experimental PMR provided that the line shape of the modulation is known, once $B_\mathrm{e}$ is determined from the analysis of the experimental trace. An alternative and more familiar way to experimentally determine $V_0$ is from the amplitude of CO\@. In the past, several groups compared $V_0$'s deduced by the two different methods for the same samples. \cite{Kato97,Soibel97,Emeleus98,Long99} In all cases, $V_0$'s deduced by PMR and by CO considerably disagree, with the former usually giving larger values. Part of the discrepancy may be attributable to underestimation of $V_0$ by CO, resulting from disregarding the proper treatment of the decay of the CO amplitude by scattering. \cite{Boggild95,Paltiel97,Endo00e} However, the most serious source of the disagreement appears to lie in the difficulty in identifying the position of $B_\mathrm{e}$ from an experimental PMR trace, which was taken, on a rather \textit{ad hoc} basis, as either the peak, \cite{Kato97,Soibel97,Emeleus98} or the position for steepest slope. \cite{Long99} It is therefore necessary to find out the rule to determine the exact position of $B_\mathrm{e}$. This is one of the purposes of the present paper. We will show below that $B_\mathrm{e}$ can be identified, when $V_0$ is small enough, as an inflection point at which the curvature of PMR changes from concave down to concave up. Another target of the present paper is the magnitude of PMR\@. The magnitude should also depend on $V_0$ as well as on other parameters of ULSL samples. The subject has been treated in theories by both numerical \cite{Menne98,Zwerschke98} and analytical \cite{Mirlin01} calculations. However, analyses of experimental PMR is so far restricted to the qualitative level \cite{Beton90P} that the magnitude increases with $V_0$. To the knowledge of the present authors, no effort has been made to date to quantitatively explain the magnitude of PMR, using the full knowledge of experimentally obtained sample parameters, $V_0$, $n_e$, the mobility $\mu$, and the quantum, or single-particle mobility $\mu_\mathrm{s}$. Such quantitative analysis has been done in the present paper for ULSL samples with relatively small periods and modulation amplitudes that allow determining reliable values of $V_0$ from the CO amplitude. \cite{Endo00e} The result demonstrates that magnetoresistance attributable to SO is much smaller than the observed PMR\@. We propose an alternative mechanism that accounts for the major part of PMR\@.
After detailing the ULSL samples used in the present study in Sec.\ \ref{smpldetail}, we delineate in Sec.\ \ref{SOcalc} a simple analytic formula to be used to estimate the contribution of SO to PMR\@. Experimentally obtained PMR traces are presented and compared to the estimated SO-contribution in Sec.\ \ref{exppmr}, leading to the introduction of another mechanism, the contribution from drift velocity of incompleted cyclotron orbits, in Sec.\ \ref{driftvel}, which we believe dominates the PMR for our present ULSL samples. Some discussion is given in Sec.\ \ref{discussion}, followed by concluding remarks in Sec.\ \ref{conclusion}.

\section{Characteristics of samples \label{smpldetail}}
\begin{table}
\caption{List of samples\label{Sampletbl}}
\begin{ruledtabular}
\begin{tabular}{ccccccc}
No. & $a$ (nm) & Hall-bar size ($\mu$m$^{2}$) & back gate \\
\hline
1 & 184 & 64$\times$37 & $\times$ \\
2 & 184 & 64$\times$37 & $\times$ \\
3 & 161 & 44$\times$16 & $\bigcirc$ \\
4 & 138 & 44$\times$16 & $\bigcirc$ \\
\end{tabular}
\end{ruledtabular}
\end{table}

\begin{figure}[tb]
\includegraphics[bbllx=50,bblly=20,bburx=780,bbury=550,width=8.5cm]{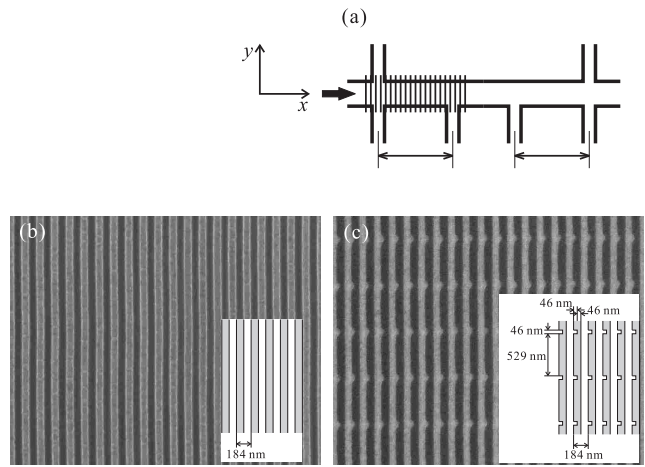}%
\caption{(a) Schematic drawing of the sample with voltage probes for measuring modulated (ULSL) and unmodulated (reference) part. (b),(c) Scanning electron micrographs of the EB-resist gratings that introduce strain-induced potential modulation. Darker areas correspond to the resist. A standard line-and-space pattern (b) was utilized for samples 1, 3 and 4. Sample 2 employed a patterned grating (c) designed to partially relax the strain. \label{samples}}
\end{figure}

We examined four ULSL samples with differing periods $a$, as tabulated in Table \ref{Sampletbl}. The samples were prepared from the same GaAs/AlGaAs single-heterostructure 2DEG wafer with the heterointerface residing at the depth $d$=90 nm from the surface, and having Al$_{0.3}$Ga$_{0.7}$As spacer layer thickness of $d_\mathrm{s}$=40 nm. A grating of negative electron-beam (EB) resist placed on the surface introduced potential modulation at the 2DEG plane through strain-induced piezoelectric effect. \cite{Skuras97} To maximize the effect, the direction of modulation ($x$-direction) was chosen to be along a $<$110$>$ direction. For a fixed crystallographic direction, the amplitude of the strain-induced modulation is mainly determined by the ratio $a/d$. Figures \ref{samples}(b) and (c) display scanning electron micrographs of the gratings. Samples 1, 3, and 4 utilized a simple line-and-space pattern as shown in (b). For sample 2, we employed a patterned grating depicted in (c); the ``line'' of resist was periodically notched in every 575 nm by width 46 nm. The width was intended to be small enough (much smaller than $d$) so that the notches introduce only negligibly small modulation themselves but act to partially relax the strain. The use of the patterned grating enabled us to attain smaller $V_0$ than sample 1, which has the same period $a$=184 nm. As shown in Fig.\ \ref{samples}(a), we used Hall bars with sets of voltage probes that enabled us to measure the section with the grating (ULSL) and that without (reference) at the same time. Resistivity was measured by a standard low-frequency ac lock-in technique. Measurements were carried out at $T$=1.4 and 4.2 K, both bearing essentially the same result. We present the result for 4.2 K in the following.

\begin{figure}[bth]
\includegraphics[bbllx=20,bblly=70,bburx=430,bbury=800,width=8.5cm]{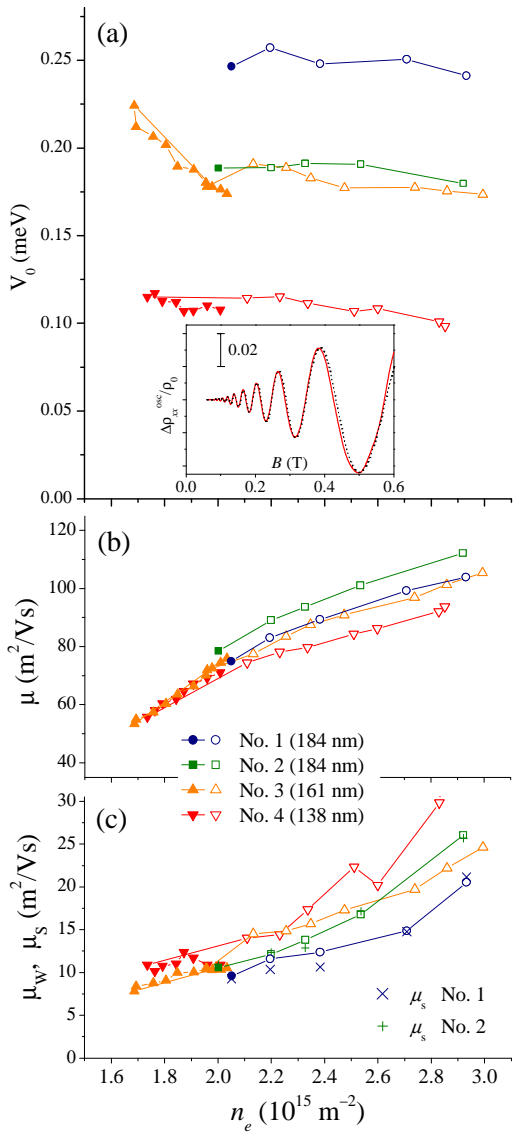}%
\caption{(Color online) Sample parameters as a function of the electron density $n_e$, varied either by LED illumination (open symbols) or by back-gate voltage (solid symbols). (a) Modulation amplitude $V_0$. (b) Mobility $\mu$. (c) Damping parameter $\mu_\mathrm{W}$ of CO\@. Quantum mobility $\mu_\mathrm{s}$ for samples 1 and 2 are also plotted by $\times$ and $+$, respectively. Inset in (a) shows $\Delta\rho_{xx}^\mathrm{osc}/\rho_0$ experimentally obtained by subtracting a slowly varying background from the magnetoresistance trace (for sample 2 at $n_e$=2.20$\times$10$^{15}$ m$^{-2}$, shown by solid trace) and calculated by Eq.\ (\ref{COosc}) using $V_0$ and $\mu_\mathrm{W}$ as fitting parameters (dotted trace, showing almost perfect overlap with the experimental trace). \label{properties}}
\end{figure}

To investigate the behavior of PMR under various values of sample parameters, $n_e$ was varied from about 2.0 to 3.0$\times$10$^{15}$ m$^{-2}$, employing persistent photoconductivity effect through step-by-step illumination with an infrared light-emitting diode (LED). Samples 3 and 4 were equipped with a back gate, which was also used to alter $n_e$ approximately between 1.7 and 2.0$\times$10$^{15}$ m$^{-2}$. The electron density $n_e$ was measured by the period of CO or Shubnikov-de Haas (SdH) oscillation, and also by Hall resistivity. Concomitant with the change of $n_e$, parameters associated with the random potential scattering, $\mu$ and $\mu_\mathrm{s}$, also vary. Plots of $\mu$ and $\mu_\mathrm{s}$ (the latter only for samples 1 and 2) versus $n_e$ are presented in Figs.\ \ref{properties}(b) and (c), respectively. Quantum mobility $\mu_\mathrm{s}$ is deduced from the damping of the SdH oscillation \cite{Coleridge91} of the unmodulated section of the Hall bar.

The amplitude $V_0$ of the modulation was evaluated from the amplitude of CO\@. In a previous publication, \cite{Endo00e} the present authors reported that the oscillatory part of the magnetoresistance is given, for $V_0$ much smaller than the Fermi Energy $E_\mathrm{F}$, $\eta$$\equiv$$V_0/E_\mathrm{F}$$\ll$1, by 
\begin{eqnarray}
\frac{\Delta\rho_{xx}^\mathrm{osc}}{\rho_0} &=& A\left(\frac{\pi}{\mu_\mathrm{W}B}\right)A\left(\frac{T}{T_a}\right) \nonumber \\
 & &\!\!\!\!\!\! \frac{1}{2\sqrt{2\pi}}\frac{1}{\Phi_0{\mu_\mathrm{B}^*}^2}\frac{\mu^2}{a}\frac{V_0^2}{n_e^{3/2}}|B| \sin\left(2\pi \frac{2R_\mathrm{c}}{a}\right),\label{COosc}
\end{eqnarray}
where $A(x)$=$x/\sinh(x)$, $k_\mathrm{B}T_a$$\equiv$$(1/2\pi^2)(ak_\mathrm{F}/2)\hbar\omega_\mathrm{c}$ with $\omega_\mathrm{c}$=$e|B|/m^*$ the cyclotron angular frequency, $\Phi_0$=$h/e$ the flux quantum, and $\mu_\mathrm{B}^*$$\equiv$$e\hbar/2m^*$($\simeq$0.864 meV/T for GaAs, an analogue of the Bohr magneton with the electron mass replaced by the effective mass $m^*$$\simeq$0.067$m_e$). Apart from the factor $A(\pi/\mu_\mathrm{W}B)$, which governs the damping of CO by scattering, Eq.\ (\ref{COosc}) is identical to the formula calculated by first order perturbation theory. \cite{Peeters92} The parameter $\mu_\mathrm{W}$ was shown in Ref.\ \onlinecite{Endo00e} to be approximately equal to $\mu_\mathrm{s}$, in accordance with the formula given for low magnetic field in the theory by Mirlin and W\"olfle. \cite{Mirlin98} Measured $\Delta\rho_{xx}^\mathrm{osc}/\rho_0$ for the present samples are also described by Eq.\ (\ref{COosc}) very well, as exemplified in the inset of Fig.\ \ref{properties}(a). So far, we have treated the modulation as having a simple sinusoidal profile $V_0\cos(2\pi x/a)$, and have tacitly neglected the possible presence of higher harmonics. Although the Fourier transforms of $\Delta\rho_{xx}^\mathrm{osc}/\rho_0$ do reveal small fraction of the second- (and also the third- for samples 1 and 2) harmonics, \cite{Endo05HH} their smallness along with the power dependence on $V_0$ of the relevant resistivities [to be discussed later, see Eqs.\ (\ref{aprhoxx}) and (\ref{drift})] justifies neglecting them to a good approximation. The parameters $V_0$ and $\mu_\mathrm{W}$ obtained by fitting Eq.\ (\ref{COosc}) to experimental traces are plotted in Figs.\ \ref{properties}(a) and (c), respectively. The latter shows $\mu_\mathrm{W}$$\simeq$$\mu_\mathrm{s}$, confirming our previous result. $V_0$ does not depend very much on $n_e$ when $n_e$ is varied by LED illumination, but increases with decreasing $n_e$ when the back gate is used, the latter resembling a previous report. \cite{Soibel97} The dependence of $V_0$ on $n_e$ is discussed in detail elsewhere. \cite{Endo05MA} Since $a$ and $d$ are of comparable size, $V_0$ rapidly increases with the increase of $a$ (with exception, of course, of sample 2 whose amplitude is close to that of sample 3). Since 6$\le$$E_F$$\le$11 meV for the range of $n_e$ encompassed in the present study, the condition $\eta$$\ll$1 is fulfilled for all the measurements shown here ($\eta$=0.010$-$0.034).

\section{Calculation of the contribution of streaming orbits\label{SOcalc}}
\begin{figure}[tb]
\includegraphics[bbllx=10,bblly=90,bburx=590,bbury=250,width=8.5cm]{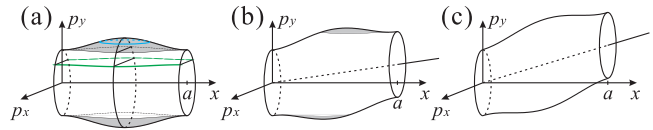}%
\caption{(Color online) Fermi surface in the $x$-$p_x$-$p_y$ space for (a) $\beta$=0, (b) 0$<$$\beta$$<$1, and (c) $\beta$=1. Each electron orbit is specified by the cross section of the Fermi surface by a constant-$p_y$ plane. Streaming orbits are present in the shaded area.\label{Fermi}}
\end{figure}
\begin{figure}
\includegraphics[bbllx=40,bblly=50,bburx=560,bbury=400,width=8.5cm]{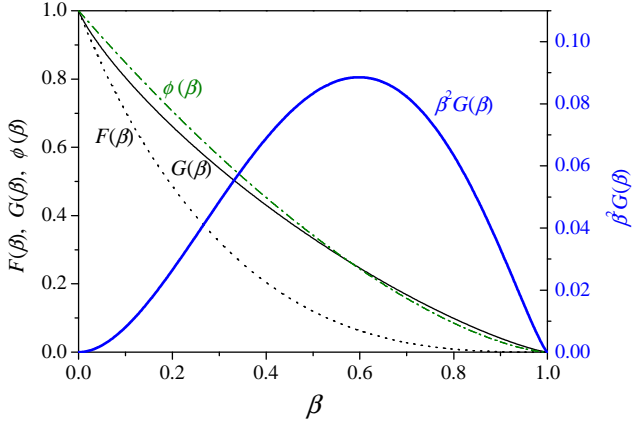}%
\caption{(Color online) Functions $F(\beta)$, $G(\beta)$, $\phi(\beta)$ (thin dotted, solid, and dash-dotted lines, respectively, left axis), and $\beta^2G(\beta)$ (thick solid line, right axis).\label{FGbeta}}
\end{figure}

In this section, we describe a simple analytic calculation for estimating the contribution of SO to magnetoresistance. The calculation is a slight modification of a theory by Matulis and Peeters, \cite{Matulis00} the theory in which semiclassical conductance was calculated for 2DEG under unidirectional magnetic field modulation with zero average. We modify the theory to the case for potential modulation $V_0 \cos(2\pi x/a)$, and extend it to include a uniform magnetic field $-B$ (the minus sign is selected just for convenience). The Hamiltonian describing the motion of electrons is given by
\begin{equation}
\varepsilon(x,p_x,p_y) = \frac{1}{2m^*}\left[p_x^2+(p_y-eBx)^2\right]+V_0\cos\left(\frac{2 \pi x}{a}\right),\label{Hamiltonian}
\end{equation}
in the Landau gauge {\bf A}=$(0,-Bx,0)$, where {\bf p}=$(p_x,p_y)$ denotes canonical momentum. Using electron velocities
\begin{equation}
v_x=\frac{\partial \varepsilon}{\partial p_x}=\frac{p_x}{m^*},\quad v_y=\frac{\partial \varepsilon}{\partial p_y}=\frac{p_y-eBx}{m^*},\label{velocity}
\end{equation}
the conductivity tensor reads (including the factor 2 for spin degeneracy)
\begin{equation}
\sigma_{ij} = \frac{2 e^2}{(2\pi \hbar)^2}\frac{1}{L_x}\int_0^{L_x}\!\!\!dx\int_{-\infty}^{\infty}\!\!\!dp_x\int_{-\infty}^{\infty}\!\!\!dp_y\tau_\mathrm{s} v_iv_j (-\frac{\partial f}{\partial \varepsilon}),\label{conductivity}
\end{equation}
where $L_x$ represents the extent of the sample in the $x$-direction and $\tau_\mathrm{s}$ denotes an appropriate scattering time to be discussed later. \cite{EquiChambers} Since the system is periodic in $x$-direction and each SO is confined in a single period, the integration over $x$, $L_x^{-1}\int_0^{L_x} dx$, can be reduced to one period, $a^{-1}\int_0^a dx$, in calculating the conductivity from SO\@. The derivative $-\partial f/\partial \varepsilon$ of the Fermi distribution function $f(\varepsilon)$=$\{1+\exp[(\varepsilon-E_\mathrm{F})/k_\mathrm{B}T]\}^{-1}$ may be approximated by the delta function $\delta(\varepsilon-E_\mathrm{F})$ at low temperatures, $T$$\ll$$E_\mathrm{F}/k_\mathrm{B}$. Therefore the problem boils down to the integration of $\tau_\mathrm{s}v_iv_j$ over relevant part of the Fermi surface $\varepsilon(x,p_x,p_y)=E_\mathrm{F}$ in the $(x,p_x,p_y)$ space. Fermi surface is depicted in Fig.\ \ref{Fermi} for three different values of $\beta$$\equiv$$B/B_\mathrm{e}$. Since the Hamiltonian Eq.\ (\ref{Hamiltonian}) does not explicitly include $y$, $p_y$ is a constant of motion that specify an orbit; an orbit is given by the cross section of the Fermi surface by a constant-$p_y$ plane. The presence of SO is indicated by the shaded area in Fig.\ \ref{Fermi}. The ratio of SO to all the orbits is maximum at $\beta$=0, decreases with increasing $\beta$, and disappears at $\beta$=1.

Before continuing the calculation, we now discuss an adequate scattering time to choose. At variance with Ref.\ \onlinecite{Matulis00}, we adopt here unweighted single-particle scattering time $\tau_\mathrm{s}$=$\mu_\mathrm{s}m^*/e$. The choice is based on the fact that the angle $\theta$=$\arctan(v_x/v_y)$ of the direction of the velocity with respect to the $y$-axis is very small for electrons belonging to SO in our ULSL samples having small $\eta$=$V_0/E_\mathrm{F}$. The maximum of $|\theta|$ at a position $u$$\equiv$$2\pi x/a$ can be approximately written as $[\eta\varphi(\beta,u)]^{1/2}$ with
\begin{equation}
\varphi(\beta,u)\equiv\sqrt{1-\beta^2}+\beta\arcsin\beta-\cos u-\beta u,
\label{varphi}
\end{equation}
whose maximum over $u$ is given by $[2\eta\phi(\beta)]^{1/2}$ with $\phi(\beta)$$\equiv$$\sqrt{1-\beta^2}+\beta\arcsin{\beta}-(\pi/2)\beta$, where $|\phi(\beta)|$$\leq$1 for $|\beta|$$\leq$1 (see Fig.\ \ref{FGbeta}). Since $|\theta|$ is much smaller than the average scattering angle $\theta_\mathrm{scat}$$\sim$$\sqrt{2\mu_\mathrm{s}/\mu}$$\simeq$0.5 rad estimated for our present 2DEG wafer, electrons are kicked out of SO by virtually any scattering event regardless of the scattering angle involved, letting $\tau_\mathrm{s}$ to be the appropriate scattering time.

The integration Eq.\ (\ref{conductivity}) over the shaded area gives the correction to the conductivity owing to SO, to the leading order in $\eta$, as
\begin{eqnarray}
\frac{\delta\sigma_{xx}^\mathrm{SO}}{\sigma_0} &=& -\frac{2}{2\pi^2}\frac{\mu_\mathrm{s}}{\mu}\int_{\arcsin\beta}^{u_1(\beta)}\frac{2}{3}[\eta\varphi(\beta,u)]^{3/2}du \nonumber \\
 &=& -\frac{32\sqrt{2}}{9\pi^2}\frac{\mu_\mathrm{s}}{\mu}\eta^{3/2}F(\beta),\label{sgmxx}
\end{eqnarray}
where the minus sign results because electrons trapped in SO cannot carry current over the (macroscopic) sample in $x$-direction and therefore should be deducted from the conductivity, and
\begin{eqnarray}
\frac{\delta\sigma_{yy}^\mathrm{SO}}{\sigma_0} &=& \frac{2}{2\pi^2}\frac{\mu_\mathrm{s}}{\mu}\int_{\arcsin\beta}^{u_1(\beta)}2[\eta\varphi(\beta,u)]^{1/2}du \nonumber \\
 &=& \frac{8\sqrt{2}}{\pi^2}\frac{\mu_\mathrm{s}}{\mu}\eta^{1/2}G(\beta),\label{sgmyy}
\end{eqnarray}
and $\delta\sigma_{xy}^\mathrm{SO}$=$\delta\sigma_{yx}^\mathrm{SO}$=0, where $\sigma_0$=$E_\mathrm{F}e^2\tau/\pi\hbar^2$ represents the Drude conductivity. The factor 2 in the first equalities accounts for the two equivalent SO areas at the upper and the lower bounds of $p_y$. The functions $F(\beta)$ and $G(\beta)$ are defined as
\begin{equation}
F(\beta)\equiv\frac{3}{16\sqrt{2}}\int_{\arcsin\beta}^{u_1(\beta)}[\varphi(\beta,u)]^{3/2}du
\label{Fbeta}
\end{equation}
and
\begin{equation}
G(\beta)\equiv\frac{1}{4\sqrt{2}}\int_{\arcsin\beta}^{u_1(\beta)}[\varphi(\beta,u)]^{1/2}du,
\label{Gbeta}
\end{equation}
where the upper limit of integration $u_1(\beta)$ is the solution of $\varphi(\beta,u_1)$=0 other than $\arcsin\beta$. Both $F(\beta)$ and $G(\beta)$ monotonically decrease from 1 to 0 while $\beta$ varies from 0 to 1, as shown in Fig.\ \ref{FGbeta}. Since $\delta\sigma_{xx}^\mathrm{SO}$$/$$\delta\sigma_{yy}^\mathrm{SO}$$\propto$$\eta$, $\delta\sigma_{xx}^\mathrm{SO}$$\ll$$\delta\sigma_{yy}^\mathrm{SO}$ for $\eta$$\ll$1. Correction to the resistivity by SO can be obtained by inverting the conductivity tensor,
\begin{eqnarray}
\frac{\delta\rho_{xx}^\mathrm{SO}}{\rho_0} &=& \left\{\frac{\delta\sigma_{xx}^\mathrm{SO}}{\sigma_0}+\left[1+(B \mu)^2\frac{\delta\sigma_{yy}^\mathrm{SO}/\sigma_0}{1+\delta\sigma_{yy}^\mathrm{SO}/\sigma_0}\right]^{-1}\right\}^{-1}\!\!\!\!\!-1 \nonumber \\
 &\simeq& -\frac{\delta\sigma_{xx}^\mathrm{SO}}{\sigma_0}+(B \mu)^2 \frac{\delta\sigma_{yy}^\mathrm{SO}}{\sigma_0} \nonumber \\
 &=& \frac{32\sqrt{2}}{9\pi^2}\frac{1}{{(\Phi_0\mu_\mathrm{B}^*)}^{3/2}}\frac{\mu_\mathrm{s}}{\mu}\frac{V_0^{3/2}}{n_e^{3/2}}F(\beta) \nonumber \\
 & & + \frac{4\sqrt{2}}{\pi}\frac{1}{{\Phi_0}^{1/2}{\mu_\mathrm{B}^*}^{5/2}}\frac{\mu_\mathrm{s}\mu}{a^2}\frac{V_0^{5/2}}{n_e^{3/2}}\beta^2G(\beta).
\end{eqnarray}
For small $\eta$, $\delta\sigma_{xx}^\mathrm{SO}$$/$$\sigma_{0}$ can be neglected and consequently
\begin{equation}
\frac{\delta\rho_{xx}^\mathrm{SO}}{\rho_0} \simeq \frac{4\sqrt{2}}{\pi}\frac{1}{{\Phi_0}^{1/2}{\mu_\mathrm{B}^*}^{5/2}}\frac{\mu_\mathrm{s}\mu}{a^2}\frac{V_0^{5/2}}{n_e^{3/2}}\beta^2G(\beta).\label{aprhoxx}
\end{equation}
The correction therefore increases in proportion to $\mu$, $\mu_\mathrm{s}$, and $V_0^{5/2}$, and decreases with $a$ and $n_e$. The function $\beta^2 G(\beta)$ is also plotted in Fig.\ \ref{FGbeta}, which takes maximum at $\beta$$\simeq$0.6 and vanishes at $\beta$=1. Our final result Eq.\ (\ref{aprhoxx}) is identical to Eq.\ (41) of Ref.\ \onlinecite{Mirlin01}, which is deduced for the case $\eta$$\ll$$\mu_\mathrm{s}/\mu$. (For larger $\eta$, Ref.\ \onlinecite{Mirlin01} gives somewhat different formula that is proportional to $V_0^{7/2}$). Note that our $\phi(\beta)$ and $G(\beta)$ are identical to the functions denoted as $\Phi_1(\beta)$ and $\Phi(\beta)$, respectively, in Ref.\ \onlinecite{Mirlin01}. In the following section, Eq.\ (\ref{aprhoxx}) will be compared with experimental traces.

\section{Positive magnetoresistance obtained by experiment\label{exppmr}}
\begin{figure}[tb]
\includegraphics[bbllx=20,bblly=30,bburx=600,bbury=710,width=8.5cm]{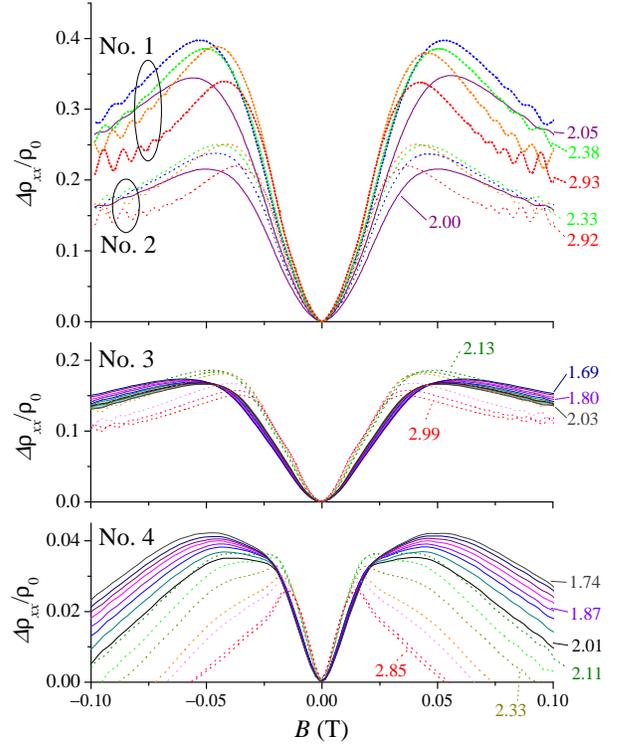}%
\caption{(Color online) Magnetoresistance traces for various values of $n_e$. Selected values of $n_e$ are noted in the figure (in 10$^{15}$ m$^{-2}$). Dotted traces indicate that the $n_e$ is attained by LED illumination. Note that the vertical scale is expanded by five times for sample 4.\label{PMRraw}}
\end{figure}

\begin{figure}[tb]
\includegraphics[bbllx=20,bblly=80,bburx=600,bbury=780,width=8.5cm]{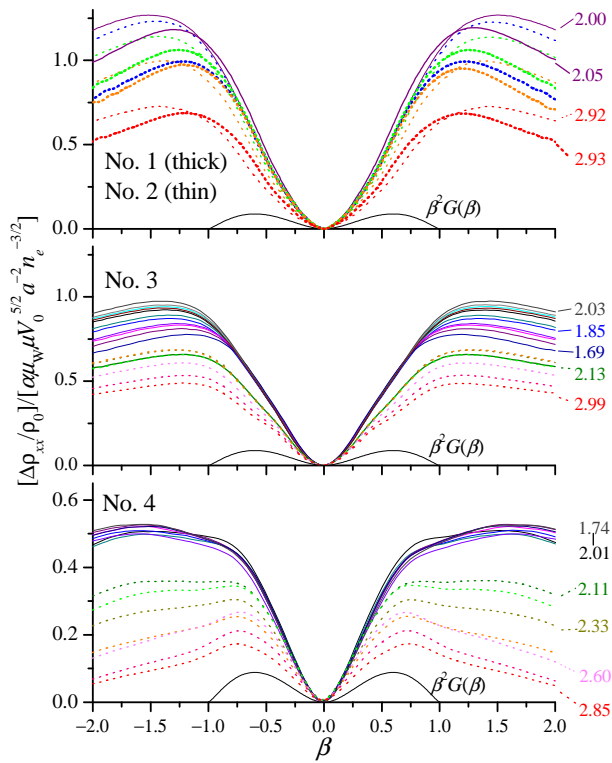}%
\caption{(Color online) Replot of Fig.\ \ref{PMRraw} with abscissa normalized by the extinction field $B_\mathrm{e}$ and ordinate by the sample-parameter-dependent prefactor in Eq.\ (\ref{aprhoxx}), $\alpha \mu_\mathrm{W} \mu V_0^{5/2} a^{-2} n_e^{-3/2}$, with the coefficient $\alpha$$\equiv$$4\sqrt{2}\pi^{-1}\Phi_0^{-1/2}{\mu_\mathrm{B}^*}^{-5/2}$$\simeq$4.04$\times$10$^7$ T$^2$meV$^{-5/2}$m$^{-1}$. Vertical scale is expanded twice for sample 4. The function $\beta^2G(\beta)$ is also plotted for comparison. \label{scale}}
\end{figure}

Figure \ref{PMRraw} shows low-field magnetoresistance traces for samples 1$-$4 for various values of $n_e$. Solid curves represent measurements before illumination ($n_e$ varied by the back gate) and dotted curves are traces for $n_e$ varied by LED illumination (back gate voltage=0 V). The magnitude of PMR shows clear tendency of being large  for samples having larger $V_0$. By contrast, the peak positions do not vary much between samples. To facilitate quantitative comparison with Eq.\ (\ref{aprhoxx}), Fig.\ \ref{PMRraw} is replotted in Fig.\ \ref{scale}, with both horizontal and vertical axes scaled with appropriate parameters: the horizontal axis is normalized by $B_\mathrm{e}$ calculated by Eq.\ (\ref{Be}) using experimentally deduced $n_e$ and $V_0$ shown in Fig.\ \ref{properties}; the vertical axis is normalized by the prefactor in Eq.\ (\ref{aprhoxx}) with $\mu_\mathrm{s}$ replaced by $\mu_\mathrm{W}$, identifying the two parameters. \cite{musmuw} Magnetoresistance owing to SO will then be represented by a universal function $\beta^2G(\beta)$, which is also plotted in the figures. It is clear from the figures that experimentally observed PMR is much larger than that calculated by Eq.\ (\ref{aprhoxx}). Furthermore, the peaks appear at $B$$>$$B_\mathrm{e}$, i.e., where SO have already disappeared, for all traces for samples 1$-$3 and traces with smaller $n_e$ for sample 4. The peak position is by no means fixed, but depends on the sample parameters. This observation argues against the interpretation of PMR being solely originating from SO\@. Rather, we interpret that SO accounts for only a small fraction of the PMR, as suggested by Fig.\ \ref{scale}, and that the rest is ascribed to another effect to be discussed in the next section. In fact, humps that appear to correspond to the component $\beta^2G(\beta)$ can readily be recognized in traces with larger $n_e$ for sample 4, superposed on a slowly increasing component of PMR\@. The humps terminate at around $|\beta|$=1, where the total PMR changes the sign of the curvature. With the increase of $\eta$($\propto$$V_0/n_e$) either by decreasing $n_e$ (upper traces for sample 4) or by increasing $V_0$ (samples 1$-$3), $\beta^2G(\beta)$ makes progressively smaller contribution to the total PMR, and becomes difficult to be distinguished from the background.

\begin{figure}
\includegraphics[bbllx=20,bblly=100,bburx=580,bbury=750,width=8.5cm]{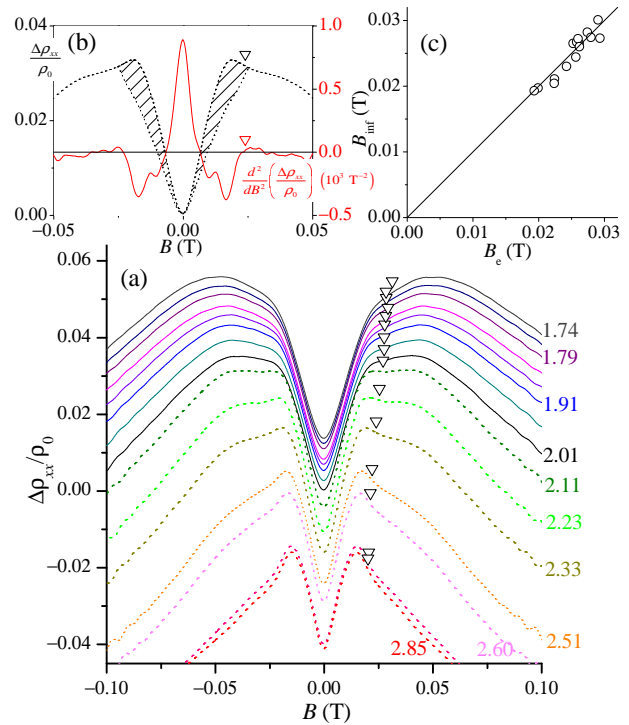}%
\caption{(Color online) (a) Magnetoresistance traces for sample 4 with the inflection point $B_\mathrm{inf}$ marked by downward open triangles. Traces are offset proportionally to the change in $n_e$. Selected values of $n_e$ in 10$^{15}$ m$^{-2}$ are noted in the figure. (b) Illustration of the procedure to pick up $B_\mathrm{inf}$ (an example for $n_e$=2.33$\times$10$^{15}$ m$^{-2}$). The point at which the second derivative $(d^2/dB^2)(\Delta\rho_{xx}/\rho_0)$ (solid curve, right axis) crosses zero upward (marked by open downward triangle) is identified as $B_\mathrm{inf}$. $B_\mathrm{inf}$ is marked also on $\Delta\rho_{xx}/\rho_0$ (dotted curve, left axis). Shaded area indicates the contribution from SO\@. (c) Plot of $B_\mathrm{inf}$ versus $B_\mathrm{e}$ calculated by Eq.\ (\ref{Be}) using experimentally obtained $V_0$. The line represents $B_\mathrm{inf}$=$B_\mathrm{e}$. \label{offset}}
\end{figure}

As has been inferred just above, the interpretation that the contribution $\delta\rho_{xx}^\mathrm{SO}/\rho_0$ from SO is superimposed on another slowly increasing background component offers an alternative way to determine $B_\mathrm{e}$: $B_\mathrm{e}$ can be identified with the end of the hump, namely, the inflection point $B_\mathrm{inf}$ where the curvature of the total PMR changes from concave down, inherited from $\beta^2G(\beta)$, to concave up. To be more specific, $B_\mathrm{inf}$ is determined as a point where the second derivative $(d^2/dB^2)(\Delta\rho_{xx}/\rho_0)$ changes sign from negative to positive as illustrated in Fig.\ \ref{offset}(b). The inflection point $B_\mathrm{inf}$ is marked by a downward open triangle both in $(d^2/dB^2)(\Delta\rho_{xx}/\rho_0)$ (solid) and $\Delta\rho_{xx}/\rho_0$ (dotted) traces. [$(d^2/dB^2)(\Delta\rho_{xx}/\rho_0)$ shows oscillatory features at low field, which are attributed to the geometric resonance of Bragg-reflected cyclotron orbits. \cite{Endo05N}] Figure \ref{offset}(a) illustrates the shift of $B_\mathrm{inf}$ with $n_e$. The plot of $B_\mathrm{inf}$ versus $B_\mathrm{e}$ shown in Fig.\ \ref{offset}(c) demonstrates that $B_\mathrm{inf}$ is actually identifiable with $B_\mathrm{e}$. Thus it is now possible to deduce reliable values of $V_0$ from PMR: by replacing $B_\mathrm{e}$ with $B_\mathrm{inf}$ in Eq.\ (\ref{Be}). Unfortunately this method is applicable only for samples with very small $\eta$. For samples 1$-$3, it is difficult to find clear inflection points because of the dominance of the slowly increasing component; $(d^2/dB^2)(\Delta\rho_{xx}/\rho_0)$ only gradually approaches zero from below. In the subsequent section, we discuss the origin of the slowly increasing background component of the PMR\@.

\section{Drift velocity of incompleted cyclotron orbits\label{driftvel}}

\begin{figure}[tb]
\includegraphics[bbllx=20,bblly=0,bburx=850,bbury=600,width=8.5cm]{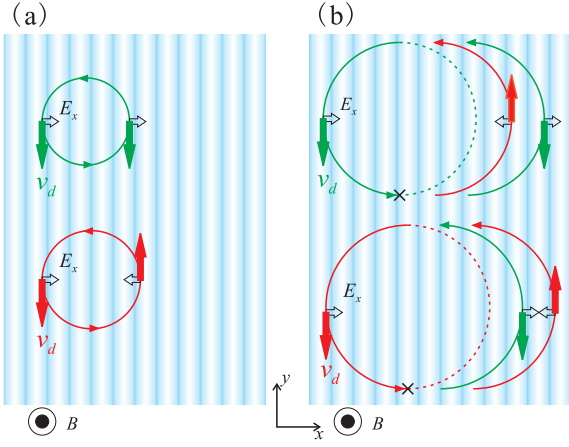}%
\caption{(Color online) Illustration of ${\bf E}\times{\bf B}$ drift velocity $v_\mathrm{d}$ affecting the electrons during the cyclotron motion. Orbits are depicted neglecting the modification due to the modulation $V(x)$=$V_0\cos(qx)$ (drifting movement and slight variation of the velocity depending on $x$) for simplicity. Top diagrams represent slightly larger $B$ than bottom ones for both (a) and (b). On averaging $v_{\mathrm{d},y}$ along an orbit, most contribution comes from minimum- and maximum-$x$ edges, as shown by solid arrows in the figure. Open arrows indicate the direction of $E_x$=$(qV_0/e)\sin(qx)$ at the edges. (a) For $B$ large enough so that electrons can travel cycles before being scattered. Depending on $B$, $v_\mathrm{d}$ at both edges are constructive (top diagram, $2R_\mathrm{c}/a$=$n+1/4$ with $n$ integer) or destructive (bottom diagram, $2R_\mathrm{c}/a$=$n-1/4$), resulting in maxima and minima in the magnetoresistance, respectively. (b) For small $B$ so that electrons are scattered before completing a cycle. The interrelation of $v_{\mathrm{d},y}$ at both edges is not simply determined by $B$. The edges affect the magnetoresistance independently.  \label{cycorb}}
\end{figure}

An important point to be noticed is that even at the low magnetic-field range $|B|$$<$$B_\mathrm{e}$ where SO is present, most of the electrons are in cyclotron-like orbits, namely the cyclotron orbits slightly modified by a weak potential modulation, as evident in Fig.\ \ref{Fermi}; SO accounts for only small fraction, order of $\eta^{1/2}$, of the whole orbits. Therefore, the contribution of these cyclotron-like orbits to the magnetoresistance should be taken into consideration in interpreting the PMR\@. We will show below that the slowly varying component of the PMR is attributable to the ${\bf E}\times{\bf B}$ drift velocity of the electrons in the cyclotron-like orbits that are scattered before completing a cycle.

It is well established that the ${\bf E}\times{\bf B}$ drift velocity resulting from the gradient of the modulation potential $\bf{E}$=$-\nabla V/(-e)$ and the applied magnetic field ${\bf B}$=$(0,0,B)$ is the origin of the CO\@. \cite{Beenakker89} For unidirectional modulation $V(x)$=$V_0\cos(qx)$ with $q$=$2\pi /a$, the drift velocity ${\bf v}_\mathrm{d}$=$({\bf E}\times{\bf B})/B^2$ has only the $y$-component,
\begin{equation}
v_{\mathrm{d},y}=\frac{qV_0}{eB}\sin(qx).\label{vdy}
\end{equation}
Electrons acquire $v_{\mathrm{d},y}$ during the course of a cyclotron revolution, whose sign alternates rapidly except for when electrons are traveling nearly parallel to the modulation ($\theta$$\simeq$0, $\pi$), i.e., around either the rightmost (maximum-$x$) or the leftmost (minimum-$x$) edges. Therefore, the contribution of the drift velocity to the conductivity comes almost exclusively from the two edges as depicted in Fig.\ \ref{cycorb} (a), which is actually experimentally verified in Ref. \onlinecite{Endo00H}. The CO is the result of the alternating occurrence by sweeping the magnetic field of the constructive and destructive addition of the effects from the two edges, as illustrated by the top and the bottom cyclotron orbits in Fig.\ \ref{cycorb} (a), respectively. With the decrease of the magnetic field, cyclotron radius $R_\mathrm{c}$ increases and consequently the probability of electrons being scattered before reaching from one to the other edge increases. As a result, the distinction between the constructive and destructive cases are blurred, letting the CO amplitude diminish more rapidly than predicted by the theories \cite{Beenakker89,Peeters92} neglecting such scattering.

The absence of CO at lower magnetic fields signifies that electrons are mostly scattered before traveling to the other edge. Although the correlation of the local drift velocities at the both edges is lost at such magnetic fields (Fig.\ \ref{cycorb} (b)), each edge can independently contribute to the conductivity. It is to this effect that we ascribe the major part of PMR in our ULSL samples. Note that the onset of CO basically coincide with the end of the PMR, bolstering this interpretation.

It can be shown, by an approximate analytic treatment of the Boltzmann's equation, that the effect actually gives rise to PMR with right order of magnitude to explain the experimentally observed slowly-varying component. For this purpose, we make use of Chambers' formula, \cite{ChambersR69,ChambersR80,Gerhardts96} representing the relaxation time approximation of Boltzmann's equation, to obtain, from the drift velocity, the component $D_{yy}$ of the diffusion tensor, 
\begin{equation}
D_{yy}=\int_{0}^{\infty} e^{-t/\tau}\langle v_{\mathrm{d},y}(t)v_{\mathrm{d},y}(0)\rangle dt,
\label{Dyy}
\end{equation}
where $\langle ... \rangle$ signifies averaging over all possible initial conditions for the motion of electrons along the trajectories. Einstein's relation is then used to obtain corresponding increment in the conductivity, $\delta\sigma_{yy}$=$e^2D(E_\mathrm{F})D_{yy}$ with $D(E_\mathrm{F})$=$m^*/\pi \hbar^2$=$(\Phi_0 \mu_\mathrm{B})^{-1}$ the density of states, and finally it is translated to the resistivity by tensor inversion, $\delta\rho_{xx}/\rho_0$=$(\omega_\mathrm{c}\tau)^2\delta\sigma_{yy}/\sigma_0$. We use unperturbed cyclotron trajectory, $x$=$X+R_\mathrm{c}\cos\theta$, for simplicity, neglecting the modification of the orbit by the modulation (and accordingly, SO is neglected in this treatment), which is justified for small $\eta$. Since the initial condition can be specified by the guiding center position $X$ and the initial angle $\theta_0$, we can write 
\begin{eqnarray}
\langle v_{\mathrm{d},y}(t)v_{\mathrm{d},y}(0)\rangle = \left( \frac{qV_0}{eB} \right)^2 \frac{1}{a}\int_{0}^{a}\!\!\!\! dX\frac{1}{2\pi}\int_{-\pi}^{\pi}\!\!\!\! d\theta_0  \hspace{2mm} \hspace{0.11\columnwidth} \nonumber \\
\sin\{ q[X+R_\mathrm{c}\cos(\theta_0+\omega_\mathrm{c}t)] \}\sin[q(X+R_\mathrm{c}\cos{\theta_0})].
\label{vyvy}
\end{eqnarray}
Therefore Eq.\ (\ref{Dyy}) can be rewritten, performing the integration over $t$ first, as
\begin{equation}
D_{yy}=\left( \frac{qV_0}{eB} \right)^2 \frac{1}{a}\int_{0}^{a}\!\!\!\! dX\frac{1}{2\pi}\int_{-\pi}^{\pi}\!\!\!\! d\theta_0 \sin[q(X+R_\mathrm{c}\cos{\theta_0})] I(\theta_0),
\label{DyyI}
\end{equation}
with
\begin{equation}
I(\theta_0)=\int_{0}^{\infty} e^{-t/\tau}\sin\{q[X+R_\mathrm{c}\cos(\theta_0+\omega_\mathrm{c}t)]\}dt.
\label{timeint}
\end{equation}

Evaluation of Eq.\ (\ref{DyyI}) for a large enough magnetic field reproduces basic features of Eq.\ (\ref{COosc}), as will be shown in the Appendix. Here, we proceed with an approximation for small magnetic fields. The approximation is rather crude but is sufficient for the purpose of getting a rough estimate of the order of magnitude.

Because of the exponential factor, only the time $t$$\alt$$\tau$ contributes to the integration of Eq.\ (\ref{timeint}).  Due to the rapidly oscillating nature of the $\sin\{ \}$ factor and the smallness of $\omega_\mathrm{c}t$$\alt$$\omega_\mathrm{c}\tau$, $I(\theta_0)$ takes a significant value only when $\theta_0$ resides in a narrow range slightly below $\sim$0 or $\sim$$\pi$, corresponding to the situation when electrons travels near the rightmost or the leftmost edge, respectively, within the scattering time. It turns out, by comparing with the numerical evaluation of Eq.\ (\ref{timeint}) using sample parameters for our present ULSL's, that the following approximate expressions roughly reproduce the right order of magnitude and the right oscillatory characteristics (the period and phase) of Eq.\ (\ref{timeint}) for low magnetic field ($|B|$$\alt$0.02 T):
\begin{equation}
I(\theta_0) \simeq \left\{
\begin{array}{ll}
\tau \pi [\sin(qX) J_0(qR_\mathrm{c})+\cos(qX) {\bf H}_0(qR_\mathrm{c})] & (\theta_0 \sim 0) \\
\tau \pi [\sin(qX) J_0(qR_\mathrm{c})-\cos(qX) {\bf H}_0(qR_\mathrm{c})] & (\theta_0 \sim \pi)
\end{array}
\right.,
\label{apprI}
\end{equation}
where $J_0(x)$ and ${\bf H}_0(x)$ represent 0-th order Bessel and Struve functions of the first kind, respectively. The approximation can be obtained by replacing the exponential factor by a constant $\omega_\mathrm{c}\tau$ and limiting the range of the time integral to include only one edge. Here we noted that the integration of $\cos(qR_\mathrm{c}\cos\theta)$ and $\sin(qR_\mathrm{c}\cos\theta)$ over the range of $\theta$ including either of the rightmost ($\theta$=0) or the leftmost ($\theta$=$\pi$) edge can be approximated (since only the close vicinity of the edges makes significant contribution to the integration) by,
\begin{equation}
\int_\mathrm{rightmost}\!\!\!\!\!\! d\theta \simeq \int_{-\pi/2}^{\pi/2}\!\!\! d\theta,\hspace{3mm} \int_\mathrm{leftmost}\!\!\!\!\!\! d\theta \simeq \int_{\pi/2}^{3\pi/2}\!\!\! d\theta,
\end{equation}
and used the relations
\begin{eqnarray}
\int_{-\pi/2}^{\pi/2}\!\!\!\!\!\!\!\! \cos(qR_\mathrm{c}\cos\theta)d\theta=\int_{\pi/2}^{3\pi/2}\!\!\!\!\!\!\!\! \cos(qR_\mathrm{c}\cos\theta)d\theta=\pi J_0(qR_\mathrm{c}) \nonumber \\
\mathrm{and}\hspace{0.8 \columnwidth} \nonumber \\
\int_{-\pi/2}^{\pi/2}\!\!\!\!\!\!\!\! \sin(qR_\mathrm{c}\cos\theta)d\theta=-\int_{\pi/2}^{3\pi/2}\!\!\!\!\!\!\!\! \sin(qR_\mathrm{c}\cos\theta)d\theta=\pi {\bf H}_0(qR_\mathrm{c}). \nonumber \\
\label{BesselStruve}
\end{eqnarray}
Substituting Eq.\ (\ref{apprI}) to Eq.\ (\ref{DyyI}) results in
\begin{equation}
D_{yy} \simeq \frac{\pi}{2} \tau \left(\frac{qV_0}{eB}\right)^2[J_0^2(qR_\mathrm{c})+{\bf H}_0^2(qR_\mathrm{c})],
\label{Dyya}
\end{equation}
and with Einstein's relation one finally obtains
\begin{equation}
\frac{\delta\rho_{xx}^\mathrm{drift}}{\rho_0}=\sqrt{\frac{\pi}{2}}\frac{1}{\Phi_0{\mu_\mathrm{B}^*}^2}\frac{\mu^2}{a}\frac{V_0^2}{n_e^{3/2}}|B|.
\label{drift}
\end{equation}
Here we made use of asymptotic expressions $J_0(x)\approx(2/\pi x)^{1/2}\cos(x-\pi/4)$ and ${\bf H}_0(x)$$\approx$$(2/\pi x)^{1/2}\sin(x-\pi/4)$ valid for large enough $x$ (corresponding to small enough $B$).

\begin{figure}
\includegraphics[bbllx=20,bblly=80,bburx=600,bbury=780,width=8.5cm]{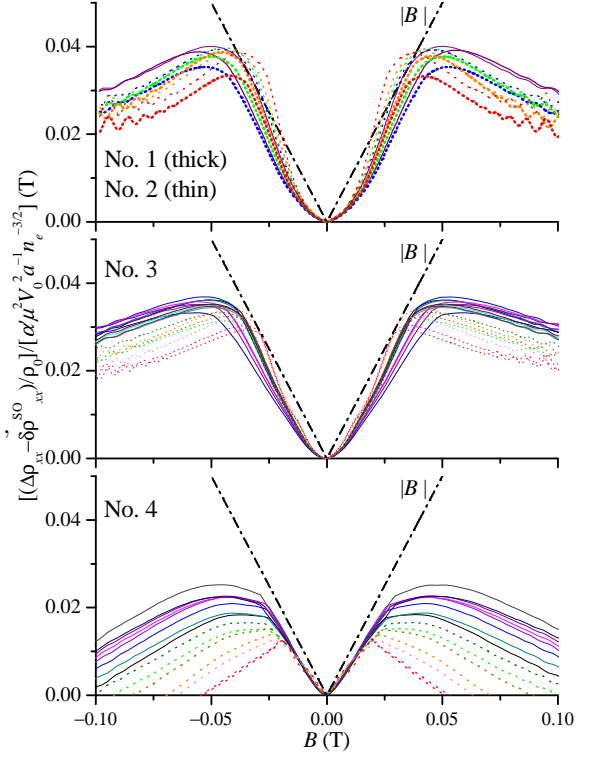}%
\caption{(Color online) Replot of magnetoresistance traces normalized by the prefactor in Eq.\ (\ref{drift}), $\alpha^\prime\mu^2V_0^2a^{-1}n_e^{-3/2}$, with $\alpha^\prime$=$(\pi/2)^{1/2}\Phi_0^{-1}{\mu_\mathrm{B}^{*}}^{-2}$=4.06$\times$10$^{14}$ T meV$^{-2}$ m$^{-2}$, after subtracting the contribution from SO, $\delta\rho_{xx}^\mathrm{SO}/\rho_0$ in Eq.\ (\ref{aprhoxx}). Contribution attributable to drift velocity of incompleted cyclotron orbits is given by $|B|$, which is also plotted by dash-dotted line. \label{dvdeCOs}}
\end{figure}

In order to compare experimentally obtained PMR with Eq.\ (\ref{drift}), PMR traces shown in Fig.\ \ref{PMRraw} are replotted in Fig.\ \ref{dvdeCOs} normalized by the prefactor in Eq.\ (\ref{drift}), after subtracting the small contribution from SO represented by Eq.\ (\ref{aprhoxx}). The scaled traces show reasonable agreement with $|B|$ at low magnetic fields, as predicted in Eq.\ (\ref{drift}), testifying that the mechanism considered here, the drift velocity from incompleted cyclotron orbits, generates PMR having the magnitude sufficient to explain the major part of PMR observed in our present ULSL samples. Possible sources of the remnant deviation, apart from the crudeness of the approximation, are (i) the neglect of higher harmonics and (ii) the neglect of negative magnetoresistance (NMR) component innate to GaAs/AlGaAs 2DEG \cite{Li03} arising from electron interactions \cite{Gornyi03,Gornyi04} or from semiclassical effect. \cite{Mirlin01N,Dmitriev02} The $n$-th harmonic gives rise to additional contribution analogous to Eq.\ (\ref{drift}) with $V_0$ and $a$ replaced by the amplitude $V_n$ of the $n$-th harmonic potential and $a/n$, respectively, and therefore, in principle, enhances the deviation. In practice, however, the effect will be small because of the small values of $V_n$ and its square dependence. On the other hand, the discrepancy can be made smaller by correcting for the NMR\@. We have actually observed NMR, which depends on $n_e$ and temperature, in the simultaneously measured ``reference'' plain 2DEG adjacent to the ULSL (see Fig.\ \ref{samples} (a)). Assuming that the NMR with the same magnitude are also present in the ULSL part and superposed on the PMR (the assumption whose validity remains uncertain at present), the correction are seen to appreciably reduce the discrepancy.

The approximation leading to Eq.\ (\ref{drift}) is valid only for very small magnetic fields. With the increase of the magnetic field, the cooperation between the leftmost and the rightmost edges is rekindled, and the magnetoresistance tends to the expression appropriate for large enough magnetic field, outlined by Eq.\ (\ref{COoscA}), which includes a non-oscillatory term (the first term) as well as the term representing CO (the second term). Note that the non-oscillatory term approaches a constant, $\alpha^\prime \mu^2 V_0^2 a^{-1} n_e^{-3/2} (m^*/2\pi e\tau)$, at small magnetic field, although the exact value of the constant is rather difficult to estimate due to the subtlety in choosing the right scattering time $\tau$, as will be discussed in the Appendix. Therefore the (linear) increase of $\delta\rho_{xx}^\mathrm{drift}/\rho_0$ with $|B|$ is expected to flatten out at a certain magnetic field. The peak in the PMR roughly marks the position of this transition, which basically corresponds to the onset of the cooperation between the two edges. Thus the peak position is mainly determined by the scattering parameters and is expected to be insensitive to $V_0$, in agreement with what has been observed in Fig.\ \ref{PMRraw}. Experimentally, the peak position $B_\mathrm{p}$ is found to be well described by an empirical formula $B_\mathrm{p}$(T)=$[4\sqrt{2\mu_\mathrm{W}(\mathrm{m}^2/\mathrm{Vs})}]^{-1}$, using $\mu_\mathrm{W}$ determined from CO\@. On the other hand, the height of the PMR peak are seen to roughly scale as ${V_0}^2$, as inferred from Fig.\ \ref{dvdeCOs}, which reveals that the normalized peak height tends to fall into roughly the same value (notably the top panel showing two samples having the same $a$ and different $V_0$), so long as the period $a$ are the same. This is better shown after correcting for the NMR effect mentioned above. The height of the normalized peak slightly decreases with decreasing $a$ (roughly proportionally to $a$), resulting in an empirical formula for the peak height $(\Delta\rho_{xx}/\rho_0)^\mathrm{peak}$ $\sim$ 3$\times$10$^{-3}$ $[\mu(\mathrm{m}^2/\mathrm{Vs})]^2$ $[V_0(\mathrm{meV})]^2$ $[n_e(10^{15}\mathrm{m}^{-2})]^{-3/2}$. (Unfortunately, sample 4 with larger $n_e$ significantly deviates from this formula.)

\section{Disscussion on the relative importance of the streaming orbit \label{discussion}}
Although PMR was thus far generally interpreted to originate from SO, contribution from mechanisms other than SO was also implied in theoretical papers. By solving Boltzmann's equation numerically, Menne and Gerhardts \cite{Menne98} calculated PMR and showed separately the contribution of SO which did not account for the entire PMR (see Fig.\ 4 in Ref.\ \onlinecite{Menne98}), leaving the rest to alternative mechanisms (although the authors did not discuss the origin futher). Mirlin \textit{et al.} \cite{Mirlin01} actually calculated contribution of drifting orbit, which is basically similar to what we have considered in the present paper. They predicted cusp-like shape for the magnetoresistance arising from this mechanism, which is not observed in experimental traces. In both papers, the major part of PMR is still ascribed to SO, with other mechanisms playing only minor roles. In the present paper, we have shown that the relative importance is the other way around in our ULSL samples. However, we would like to point out that the dominant mechanism may change with the amplitude of modulation in ULSL\@.

The reason for the contribution of SO being small in our samples can be traced back to the small amplitude of the modulation, combined with the small-angle nature of the scattering in the GaAs/AlGaAs 2DEG\@. As mentioned earlier, small $\eta$=$V_0/E_\mathrm{F}$ limits the SO within narrow angle range $|\theta|$$\leq$$\sqrt{2\eta}$, letting the electrons being scattered out of the SO even by an small-angle scattering event, hence the use of $\tau_\mathrm{s}$ in Eq.\ (\ref{conductivity}). This leads to small $\delta\sigma_{yy}^\mathrm{SO}$, since $\tau_\mathrm{s}$$\ll$$\tau$. Within the present framework, relative weight of SO in PMR decreases with increasing $\eta$, since the ratio of Eq.\ (\ref{aprhoxx}) to Eq.\ (\ref{drift}) is proportional to $\eta^{-1/2}$, in agreement with what was observed in Fig.\ \ref{scale}. However, the situation will be considerably altered with further increase in $\eta$ (typically $\eta$$\agt$0.1). Then, due to the expansion of the angle range encompassed by SO, electrons begin to be allowed to stay within SO after small-angle scattering, requiring $\tau_\mathrm{s}$ in Eq.\ (\ref{conductivity}) to be replaced by larger (possibly $B$-dependent) values. In the limit that the range of $|\theta|$ is much larger than the average scattering angle, $\tau_\mathrm{s}$ should be supplanted by ordinary transport lifetime (momentum-relaxation time) $\tau$, resulting in much larger $\delta\sigma_{yy}^\mathrm{SO}$. This largely enhances the relative importance of SO, possibly to an extent to exceed the contribution from the drift velocity. We presume that the contribution of SO is much larger than in our case in most of the experiments reported so far which showed the shift of PMR peak position with the modulation amplitude \cite{Beton90P,Kato97,Soibel97,Emeleus98,Long99}. Even in such situation, however, it will not be easy to obtain simple relation between the peak position $B_\mathrm{p}$ and the amplitude $V_0$ because of the complication by the remnant contribution from the drift velocity. In most experiments, $V_0$ is varied by the gate bias, which concomitantly alters the electron density and scattering parameters, thereby affecting the both contributions as well.

\section{Conclusions\label{conclusion}}
The positive magnetoresistance (PMR) in unidirectional lateral superlattice (ULSL) possesses two different types of mechanisms as its origin: the streaming orbit (SO) and the drift velocity of incompleted cyclotron orbit. Although virtually only the former mechanism has hitherto been taken into consideration, we have shown that the latter mechanism account for the main part of PMR observed in our ULSL samples characterized by their small modulation amplitude. The share undertaken by SO decreases with increasing $\eta$=$V_0/E_\mathrm{F}$, insofar as $\eta$ is kept small enough for the electrons in SO to be driven out even by a small-angle scattering characteristic of GaAs/AlGaAs 2DEG; $\eta$$\leq$0.034 for our samples fulfills this requirement. In this small $\eta$ regime, the peak position of PMR is not related to the modulation amplitude $V_0$ but rather determined by scattering parameters; the peak roughly coincide with the onset of commensurability oscillation (CO) that notifies the beginning of the cooperation between the leftmost and the rightmost edges in a cyclotron revolution. The height of the peak, on the other hand, are found to be roughly proportional to $V_0^2$. For small enough $\eta$, the contribution of SO becomes distinguishable as a hump superposed on slowly-increasing component and the magnetic field that marks the end of the SO, $B_\mathrm{e}$, can be identified as an inflection point of the magnetoresistance trace where the curvature changes from concave down to concave up. The extinction field $B_\mathrm{e}$ provides an alternative method via Eq.\ (\ref{Be}) to accurately determine $V_0$. We have also argued that for samples with $\eta$ much larger than ours, typically $\eta$$\agt$0.1, the relative importance of the two mechanisms can be reversed and the PMR peak position $B_\mathrm{p}$ can depend on $V_0$, although it will be difficult to deduce a reliable value of $V_0$ from $B_\mathrm{p}$.

\begin{acknowledgments}
This work was supported by Grant-in-Aid for Scientific Research in Priority Areas ``Anomalous Quantum Materials'', Grant-in-Aid for Scientific Research (C) (15540305) and (A) (13304025), and Grant-in-Aid for COE Research (12CE2004) from Ministry of Education, Culture, Sports, Science and Technology.
\end{acknowledgments}

\appendix*
\section{Approximation for higher magnetic field}
In this appendix, we delineate the approximation of Eq.\ (\ref{Dyy}) at higher magnetic field, which leads to an expression for commensurability oscillation (CO). When the velocity $v_{\mathrm{d},y}(t)$ is a periodic function of time with period $T$, Eq.\ (\ref{Dyy}) reduces to \cite{Gerhardts96}
\begin{equation}
D_{yy}=\frac{1}{1-e^{-T/\tau}}\int_{0}^{T} e^{-t/\tau}<v_{\mathrm{d},y}(t)v_{\mathrm{d},y}(0)>dt.
\label{Dyyperiodic}
\end{equation}
Using here again the unperturbed cyclotron orbit $x$=$X+R_\mathrm{c}\cos(\omega_\mathrm{c}t)$, one obtains
\begin{eqnarray}
D_{yy}=\left( \frac{qV_0}{eB} \right)^2 \frac{1}{a}\int_{0}^{a}\!\!\!\! dX\frac{1}{2\pi}\int_{-\pi}^{\pi}\!\!\!\! d\theta_0 \sin[q(X+R_\mathrm{c}\cos{\theta_0})] \nonumber \\
\frac{1}{1-e^{-T/\tau}} I_T(\theta_0) \hspace{0.03\columnwidth} \nonumber \\
\label{DyyIT}
\end{eqnarray}
with $T$=$2\pi/\omega_\mathrm{c}$ and 
\begin{equation}
I_T(\theta_0)=\int_{0}^{T} e^{-t/\tau}\sin\{q[X+R_\mathrm{c}\cos(\theta_0+\omega_\mathrm{c}t)]\}dt.
\label{timeintT}
\end{equation}
Again because of the $\sin\{ \}$ factor, the main contribution in the integration comes from the narrow band of $t$ around $\theta_0+\omega_\mathrm{c}t$$\sim$0 (or 2$\pi$ depending on the initial angle $\theta_0$) and $\pi$. For large enough $\omega_\mathrm{c}$, the band of $t$ becomes narrow enough to allow the exponential factor $e^{-t/\tau}$ to be approximated by a constant value at $t$=$-(\theta_0-k\pi)/\omega_\mathrm{c}$ (with $k$=0,1, and 2). Thus, using the relations Eq.\ (\ref{BesselStruve}), $I_T(\theta_0)$ can be approximately written, depending on the values of $\theta_0$, as
\begin{eqnarray}
I_T^<(\theta_0)\simeq \frac{\pi}{\omega_\mathrm{c}} e^\frac{\theta_0}{\omega_\mathrm{c}\tau} [\sin(qX) J_0(qR_\mathrm{c})+\cos(qX) {\bf H}_0(qR_\mathrm{c})] \nonumber \\
+\frac{\pi}{\omega_\mathrm{c}} e^\frac{\theta_0-\pi}{\omega_\mathrm{c}\tau} [\sin(qX) J_0(qR_\mathrm{c})-\cos(qX) {\bf H}_0(qR_\mathrm{c})] \nonumber \\
\label{ITthetaminus}
\end{eqnarray}
for $-\pi^{+}$$<$$\theta_0$$<$$0^{-}$ and
\begin{eqnarray}
I_T^>(\theta_0)\simeq \frac{\pi}{\omega_\mathrm{c}} e^\frac{\theta_0-\pi}{\omega_\mathrm{c}\tau} [\sin(qX) J_0(qR_\mathrm{c})-\cos(qX) {\bf H}_0(qR_\mathrm{c})] \nonumber \\
+\frac{\pi}{\omega_\mathrm{c}} e^\frac{\theta_0-2\pi}{\omega_\mathrm{c}\tau} [\sin(qX) J_0(qR_\mathrm{c})+\cos(qX) {\bf H}_0(qR_\mathrm{c})] \nonumber \\
\label{ITthetaplus}
\end{eqnarray}
for $0^{+}$$<$$\theta_0$$<$$\pi^{-}$, where the superscripts $+$ and $-$ indicate small setbacks to avoid the region where the integration have significant value. When $\theta_0$ approaches closer to the boundary, $I_T(\theta_0)$ approaches the average of the values on both sides,
\begin{equation}
I_T(\theta_0\rightarrow 0)\rightarrow [I_T^<(\theta_0)+I_T^>(\theta_0)]/2
\label{ITzero}
\end{equation}
and
\begin{equation}
I_T(\theta_0\rightarrow \pi)\rightarrow [I_T^>(\theta_0)+I_T^<(\theta_0-2\pi)]/2,
\label{ITpi}
\end{equation}
which can be shown by using the relations,
\begin{equation}
\int_{-\pi/2}^{0}\!\!\!\!\!\!\!\! \cos(qR_\mathrm{c}\cos\theta)d\theta=\int_{0}^{\pi/2}\!\!\!\!\!\!\!\! \cos(qR_\mathrm{c}\cos\theta)d\theta=\frac{\pi}{2} J_0(qR_\mathrm{c})
\label{halfBesselStruve}
\end{equation}
and other related equations corresponding to the halves of Eq.\ (\ref{BesselStruve}).
In the integration by $\theta_0$ in Eq.\ (\ref{DyyIT}), only $\theta_0$$\sim$0 and $\pi$ contributes to the integral for the same reason as before. The integration, after slightly shifting the limits of the integral from $\int_{-\pi}^\pi$ to $\int_{-\pi^+}^{\pi^+}$, yields
\begin{eqnarray}
\frac{\pi}{2\omega_\mathrm{c}} \hspace{0.8\columnwidth} \nonumber \\
\left\{ (1+e^{-\frac{2\pi}{\omega_\mathrm{c}\tau}}) \left[ \sin^2(qX) {J_0}^2(qR_\mathrm{c})+\cos^2(qX) {{\bf H}_0}^2(qR_\mathrm{c}) \right] \right. \nonumber \\
\left. +2 e^{-\frac{\pi}{\omega_\mathrm{c}\tau}} \left[ \sin^2(qX) {J_0}^2(qR_\mathrm{c})-\cos^2(qX) {{\bf H}_0}^2(qR_\mathrm{c}) \right] \right\}. \nonumber \\
\label{thetaint}
\end{eqnarray}
Finally, by averaging over $X$, Eq.\ (\ref{DyyIT}) becomes
\begin{eqnarray}
D_{yy}&=&\frac{1}{2}\left( \frac{qV_0}{eB} \right)^2
\frac{\pi}{2\omega_\mathrm{c}} \nonumber \\
&&\left\{ \coth \left(\frac{\pi}{\omega_\mathrm{c}\tau} \right) \left[ {J_0}^2(qR_\mathrm{c})+ {{\bf H}_0}^2(qR_\mathrm{c}) \right] \right. \nonumber \\
&&\left. +\sinh^{-1} \left(\frac{\pi}{\omega_\mathrm{c}\tau} \right) \left[ {J_0}^2(qR_\mathrm{c})- {{\bf H}_0}^2(qR_\mathrm{c}) \right] \right\} \nonumber \\
&\simeq& \frac{\tau}{2} \left( \frac{qV_0}{eB} \right)^2 \frac{1}{\pi q R_\mathrm{c}} \nonumber \\
&&\left[ \frac{\pi}{\omega_\mathrm{c}\tau}\coth \left(\frac{\pi}{\omega_\mathrm{c}\tau} \right) + A\left( \frac{\pi}{\omega_\mathrm{c}\tau} \right) \sin(2qR_\mathrm{c}) \right], \nonumber \\
&&\label{DyyCO}
\end{eqnarray}
which can be translated to magnetoresistance with the use of Einstein's relation, resulting in,
\begin{eqnarray}
\frac{\Delta\rho_{xx}}{\rho_0}&=&\frac{1}{2\sqrt{2\pi}}\frac{1}{\Phi_0{\mu_\mathrm{B}^*}^2}\frac{\mu^2}{a}\frac{V_0^2}{n_e^{3/2}}|B| \hspace{0.2\columnwidth} \nonumber \\
&&\left[ \frac{\pi}{\omega_\mathrm{c}\tau}\coth \left(\frac{\pi}{\omega_\mathrm{c}\tau} \right) + A\left( \frac{\pi}{\omega_\mathrm{c}\tau} \right) \sin(2qR_\mathrm{c}) \right]. \nonumber \\
&&\label{COoscA}
\end{eqnarray}
The formula agree with the asymptotic expression of Eq.\ (21) in Ref.\ \onlinecite{Mirlin98} for large enough magnetic fields. The second term in Eq.\ (\ref{COoscA}) represents CO, which reproduces qualitative features of Eq.\ (\ref{COosc}). It should be noted, however, that the anisotropic nature of the scattering in GaAs/AlGaAs 2DEG cannot be correctly treated in the present simple relaxation-time-approximation approach employing only one scattering time. The anisotropic scattering plays an important role in CO because of its high sensitivity to the small-angle scattering. Therefore Eq.\ (\ref{COoscA}) is of limited validity to describe CO in our ULSL\@. However, it is interesting to point out that if we are allowed to replace $\tau$ \textit{only} in the scattering damping factor $A(\pi/\omega_\mathrm{c}\tau)$ with the single-particle $\tau_\mathrm{s}$=$\mu_\mathrm{s}m^*/e$$\simeq$$\mu_\mathrm{W}m^*/e$, we acquire Eq.\ (\ref{COosc}) except for the thermal damping factor. The choice of $\tau_\mathrm{s}$ for the damping is not unreasonable, considering that the factor $A(\pi/\omega_\mathrm{c}\tau)$ stems from the exponential factor in Eq.\ (\ref{thetaint}), which describe the cooperativeness between the leftmost and rightmost edges that is susceptible to a small-angle scattering. [In general, velocity-velocity correlation in Eq.\ (\ref{Dyy}) for $t$$\sim$$kT/2$ with $k$=1,2,3,..., namely between the edges separated by more than half revolution of the cyclotron orbit, requires precise positioning after the revolution, which is ruined by small angle scattering. Therefore the use of $\tau_\mathrm{s}$ is reasonable. However, this does not justify the replacement \textit{only} in the damping factor.] The thermal damping factor $A(T/T_a)$ can readily be incorporated by allowing for thermal smearing of the Fermi edge, namely by taking the average over the energy of the $\sin (2qR_\mathrm{c})$ term weighted by the factor $(-\partial f/\partial \varepsilon)$.

\bibliography{lsls,magmod,ourpps,misc,negmr,notepmr}

\begin{thebibliography}{43}
\bibitem[{Equ()}]{EquiChambers}
\bibinfo{note}{It can readily be shown that Eq.\ (\ref{conductivity}) is
  equivalent to the Chambers' formula, Eq.\ (\ref{Dyy}), in the limit
  $\omega_\mathrm{c}\tau$$\ll$1, after allowing for the energy spread at a
  finite temperature.}
\bibitem[{mus()}]{musmuw}
\bibinfo{note}{From our experience, it was easier to deduce an accurate value
  of $\mu_\mathrm{W}$ from CO than to obtain $\mu_\mathrm{S}$ exactly from SdH
  oscillation, the latter readily being made inaccurate by small inhomogeneity
  in $n_e$, see Ref.\ \onlinecite{Coleridge91}.}
\end{thebibliography}

\end{document}